# Transmission resonances in above-barrier reflection of ultra-cold atoms by the Rosen-Morse potential

H.A. Ishkhanyan[1], V.P. Krainov[1], and A.M. Ishkhanyan[2]

[1]*Moscow Institute of Physics and Technology, 141700 Dolgoprudny, Russia*
[2]*Institute for Physical Research NAS of Armenia, 0203 Ashtarak-2, Armenia*

**Abstract.** Quantum above-barrier reflection of ultra-cold atoms by the Rosen-Morse potential is analytically considered within the mean field Gross-Pitaevskii approximation. Reformulating the problem of reflectionless transmission as a quasi-linear eigenvalue problem for the potential depth, an approximation for the specific height of the potential that supports reflectionless transmission of the incoming matter wave is derived via modification of the Rayleigh-Schrödinger time-independent perturbation theory. The approximation provides highly accurate description of the resonance position for all the resonance orders if the nonlinearity parameter is small compared with the incoming particle's chemical potential. Notably, the result for the first transmission resonance turns out to be exact, i.e., the derived formula for the resonant potential height gives the exact value of the first nonlinear resonance's position for all the allowed variation range of the involved parameters, the nonlinearity parameter and chemical potential. This has been shown by constructing the exact solution of the problem for the first resonance. Furthermore, the presented approximation reveals that, in contrast to the linear case, in the nonlinear case reflectionless transmission may occur not only for potential wells but also for potential barriers with positive potential height. It also shows that the nonlinear shift of the resonance position from the position of the corresponding linear resonance is approximately described as a linear function of the resonance order. Finally, a compact (yet, highly accurate) analytic formula for the $n$ th order resonance position is constructed via combination of analytical and numerical methods.

Bose-Einstein condensates of ultracold gases [1,2] provide an ideal ground for testing of many important nonlinear phenomena occurring in many-body quantum systems. An increasing number of such phenomena, e.g., the creation of topological structures such as vortices [3], the generation of bright [4] and dark [5] solitons, the self-trapping effect [6], etc., has been recently extensively studied both theoretically and experimentally. One of such important phenomena offered by the Bose-condensates is the *macroscopic* quantum tunneling through and reflection from a potential barrier (well) of a many-body wave function. This is because the basic concepts of tunneling through a barrier and above-barrier reflection of a particle are fundamental effects in quantum mechanics not present in classical physics [7]. Since the many-body macroscopic tunneling and reflection are essentially nonlinear processes, and, hence, provide a distinctly different test of the validity of quantum mechanics which is not possible for the one-particle case, these phenomena attracted a considerable attention during the last years since the experimental realization of the Bose-Einstein condensation in dilute gases of ultra-cold neutral atoms [2]. Several physical situations have been discussed including the step-like potential [8], single, finite-number and periodic (Kronig-Penney) rectangular potential barriers or wells of finite or infinite depth [9], single- and double-delta-function or a periodic sequence of delta potentials (delta-comb potential) [10], some other celebrated potentials such as the double Gaussian barrier, etc. [11].

In the present paper, we consider the above-barrier reflection of a Bose-Einstein condensate by the finite-height Rosen-Morse potential

$$V(x) = V_0 \operatorname{sech}^2(x/x_0), \qquad (1)$$

for which the solution of the corresponding linear problem is known [7]. This problem has already been addressed in several publications (see, e.g., [12-14]). In particular, the exact solution of the problem has been reported for a certain depth of the potential [13] and the reflection coefficient has been calculated for the first reflectionless transmission resonance for the case of small nonlinearity [14]. Here we focus on the reflectionless transmission resonances viewed in terms of incoming and outgoing waves.

A convenient framework for investigation of the tunnelling phenomena with Bose-Einstein condensates is the mean-field approximation described by the Gross-Pitaevskii equation [15]. This is a basic version of the nonlinear Schrödinger equation encountered in many physical systems besides Bose-condensates including, e.g., nonlinear optics, spin waves in magnetic films, Langmuir waves in hot plasmas and gravity surface waves in fluids and thus is of general physical interest. In the one-dimensional case the Gross-Pitaevskii equation is written as

$$i\hbar \frac{\partial \Psi}{\partial t} = -\frac{\hbar^2}{2m}\frac{\partial^2 \Psi}{\partial x^2} + (V(x) + g|\Psi|^2)\Psi = 0, \qquad (2)$$

where the nonlinearity parameter $g$ characterizes the mean-field self-interaction energy. Applying the ansatz $\Psi(x,t) = \exp(-i\mu t/\hbar)\psi(x)$, where $\mu$ is the chemical potential, Eq. (2) is reduced to the following time-independent version

$$-\frac{1}{2}\frac{d^2\psi}{dx^2} + (-\mu + V(x) + g|\psi|^2)\psi = 0, \qquad (3)$$

where we have adopted the system of units where the reduced Planck constant, the involved mass of the quantum particle as well as the space-scale $x_0$ are put equal to unity.

For a sufficiently large chemical potential $\mu$, since the Rosen-Morse potential vanishes at $x \to -\infty$, in the vicinity of $x = -\infty$ Eq. (3) allows a solution with asymptotic behavior $\psi \sim c_0 e^{ikx}$, $c_0 = \text{const}$, corresponding to a traveling-wave running from left to right. Here, the wave number $k$, defined as $k = \sqrt{2(\mu - |c_0|^2 g)}$, is supposed to be real so that we assume $\mu - |c_0|^2 g > 0$. Since the normalization of the wave function can always be incorporated in the definition of the nonlinearity parameter $g$, without loss of generality, we here adopt the normalization $|\psi(-\infty)|^2 = 1$, hence, below we put $|c_0| = 1$. [Note that the wave-number $k$ should be imaginary if $\mu < |c_0|^2 g$ and $g > 0$, i.e., the asymptote of the wave at $x = \pm \infty$ in this case should be an exponentially vanishing real function, an evanescent wave. Since such an asymptote describes a quantum state localized in a finite space-region (effectively, a close region embracing the potential), i.e., a bound state, it is understood that the nonlinearity modifies the condition for trapping the particle, namely, the particle is trapped if the chemical potential $\mu < |c_0|^2 g$, $g > 0$. This is, of course, an expression of the self-trapping effect [6].]

Thus, we are interested in the solution of Eq. (3) of the form

$$\psi = e^{ikx} u(x), \qquad (4)$$

where
$$k = \sqrt{2(\mu - g)}, \qquad (5)$$

and the function $u(x)$ obeys the initial condition $|u(-\infty)| = 1$. For real $k$, transformation (4) leads to the following equation for $u(x)$:

$$u_{xx} + 2ik\, u_x - [2V_0 \operatorname{sech}^2 x + 2g(|u|^2 - 1)]u = 0. \qquad (6)$$

Applying now the change of the independent variable $z = (1 + \operatorname{th} x)/2$, $z \in [0,1]$, so that $d/dx = -2z(z-1)d/dz$ and $\operatorname{sech}^2 x = -4z(z-1)$, this equation is rewritten as

$$u_{zz} + \left(\frac{1+ik}{z} + \frac{1-ik}{z-1}\right) u_z + \frac{2V_0}{z(z-1)} u - \frac{g(|u|^2 - 1)}{2z^2(z-1)^2} u = 0. \qquad (7)$$

The linear part of the obtained equation is the hypergeometric equation [16] the solution of which satisfying the initial condition discussed here is

$$u = {}_2F_1(\alpha, 1-\alpha; 1+ik; z), \qquad (8)$$

where
$$\alpha = \frac{1}{2} + \sqrt{\frac{1}{4} - 2V_0}. \qquad (9)$$

Note now that the reflectionless transmission is achieved if the boundary condition $|u(+\infty)|=1$ is satisfied. For the linear case $g=0$, when the function (8) determines the exact solution of the problem, this condition is fulfilled if $\alpha$ is a negative integer, i.e., when $\alpha=-n$, $n=1,2,3,...$ The corresponding transmission resonances are then achieved for

$$V_{Ln} = -\frac{1}{2}n(n+1). \tag{10}$$

As it is immediately seen, the resonances occur for negative $V_{Ln}$ or, in other words, the reflectionless transmission in the linear case is possible only for potential wells.

Consider now the reflectionless transmission in the nonlinear case $g \neq 0$. Note that application of the boundary condition $|u(+\infty)|=1$ defines a quasi-linear eigenvalue problem for the potential depth $V_0$ that we formulate in the following operator form

$$\hat{H}_L u + V_0 u = gF(u), \tag{11}$$

$$|u(-\infty)| = |u(+\infty)| = 1, \tag{12}$$

where $\hat{H}_L$ stands for the linear operator

$$\hat{H}_L = \frac{z(z-1)}{2}\left[\frac{d^2}{dz^2} + \left(\frac{1+ik}{z} + \frac{1-ik}{z-1}\right)\frac{d}{dz}\right] \tag{13}$$

and $F$ denotes the nonlinear function

$$F(u) = \frac{|u|^2 - 1}{4z(z-1)}u. \tag{14}$$

Since the solution to the linear problem is known [the eigenfunctions and eigenvalues are given by Eqs. (8) and (10), respectively] it is straightforward to apply the Rayleigh-Schrödinger perturbation theory to construct the approximate solution to the nonlinear problem for small $g$. Thus, we suppose that $g$ is small enough and for the $n$ th order nonlinear transmission resonance apply the expansion

$$u = u_{NLn} = u_{Ln} + gu_1 + g^2 u_2 + ..., \tag{15}$$

$$V_0 = V_{NLn} = V_{Ln} + gV_1 + g^2 V_2 + ..., \tag{16}$$

where, according to Eqs. (8), (9) and (10):

$$u_{Ln} = {}_2F_1(-n, 1+n, 1+ik, z). \tag{17}$$

Eq. (11) then reads

$$\hat{H}_L(u_{Ln} + gu_1 + ...) + (V_{Ln} + gV_1 + ...)(u_{Ln} + gu_1 + ...) = gF(u_{Ln} + gu_1 + ...) \tag{18}$$

so that equating the terms at equal powers of $g$ we obtain

$$\hat{H}_L u_{Ln} + V_{Ln} u_{Ln} = 0, \quad (19)$$

$$\hat{H}_L u_1 + (V_{Ln} u_1 + V_1 u_{Ln}) + F(u_{Ln}) = 0. \quad (20)$$

Eq. (19) is of course automatically satisfied. As regards equation (20) for $u_1$, we apply an expansion in terms of functions $u_{Ln}$:

$$u_1 = \sum_{m=1}^{\infty} a_m u_{Lm}, \quad a_m = \text{const}. \quad (21)$$

Note now that the functions $u_{Ln}$ are orthogonal in the interval $z \in [0,1]$ with weight function $z^{ik}(1-z)^{-ik}$:

$$\int_0^1 z^{ik}(1-z)^{-ik} u_{Lm} u_{Ln} \, dz = C_n \delta_{mn}, \quad (22)$$

where $\delta_{mn}$ is the Kronecker delta and $C_n$ is a constant. Substituting then expansion (21) into Eq. (20), multiplying the resultant equation by $z^{ik}(1-z)^{-ik} u_{Ln}$ and integrating over the interval $z \in [0,1]$ we immediately obtain

$$V_1 = \frac{1}{C_n} \int_0^1 z^{ik}(1-z)^{-ik} F(u_{Ln}) u_{Ln} \, dz. \quad (23)$$

Numerical simulations show that this is a very good approximation as far as the nonlinearity parameter $g$ is small. More precisely, the derived formula accurately describes the depth of the potential well for all the transmission resonance orders $n \in N$ and wave vectors $k \in [0, \infty)$ if $g < 0.25\mu$ and it provides a rather good approximation up to $g \approx (0.5 \div 0.75)\mu$.

Note now that the constant $C_n$ is calculated exactly:

$$C_n = \frac{\Gamma(1+ik)^2}{2n+1} \frac{\Gamma(n+1-ik)}{\Gamma(n+1+ik)}, \quad (24)$$

where $\Gamma$ is the Euler gamma function. Furthermore, note that for an integer $n$ the function $u_{Ln}$ is a polynomial in $z$. Hence, the integral (23) can be analytically calculated for any given order $n$. Thus, we conclude that within the limits of the Rayleigh-Schrödinger perturbation theory the first-order nonlinear shift of the depth of the reflecting potential (1) can be calculated for any reflectionless transmission resonance order. For the first three resonances [for which $V_{L1} = -1$, $V_{L2} = -3$ and $V_{L3} = -6$, see Eq. (10)] the result reads

$$V_{NL1} = -1 + \frac{g}{1 + 2(\mu - g)}, \quad (25)$$

$$V_{NL2} = -3 + \frac{9g}{7}\left(\frac{1}{1+2(\mu-g)} + \frac{2}{4+2(\mu-g)}\right), \tag{26}$$

$$V_{NL3} = -6 + \frac{2g}{11}\left(\frac{9}{1+2(\mu-g)} + \frac{15}{4+2(\mu-g)} + \frac{25}{9+2(\mu-g)}\right). \tag{27}$$

As it is immediately seen, in contrast to the linear case, the resonance position in the nonlinear case depends on the chemical potential $\mu$. Note that the shift from the value of the linear resonance's potential depth is negative for attractive interaction ($g < 0$) and is positive for repulsive interaction ($g > 0$). If the nonlinearity is repulsive and strong enough, the shift in the dept of the potential may produce a positive value of the resonance potential's height. Hence, in contrast to the linear case, in the nonlinear case reflectionless transmission may occur not only for potential wells but also for potential barriers. For instance, for the first resonance this happens when $g > (1+2\mu)/3$.

A remarkable final observation is that the formula for the first resonance, Eq. (25) turns out to be exact. This can be verified trying to find an exact solution of the exact nonlinear equation (7) as a linear function of $z$. Indeed, substituting $u = 1 + a\,z$ into Eq. (7) immediately leads to the simple solution

$$u_{NL1} = 1 - \frac{2z}{1+ik} \tag{28}$$

for a potential depth $V_0$ defined by Eq. (25). Since when passing to the linear limit by tending $g \to 0$ the depth of the potential defined by this equation becomes equal to $-1$ it is understood that this solution describes the nonlinear transmission resonance $V_{NL1}$ corresponding to the first linear resonance $V_{L1} = -1$.

Though the presented development provides a quite accurate quantitative description of the reflectionless transmission resonance position for any given resonance order $n$, however, it is also understood from Eqs. (25)-(27) that for higher order resonances the formulas become rather cumbersome. For this reason, we now attempt to construct a compact (yet, highly accurate) analytic approximation for the integral (23) via combination of analytical and numerical methods.

We start with examination of the dependence of the resonance position shift $V_1 = V_{NLn} - V_{Ln}$ on the wave vector $k = \sqrt{2(\mu-g)}$ of the incident matter-wave. This dependence for the first six resonances is shown in Figs. 2a and 2b. A remarkable immediate observation suggested by the presented graphs is that for each *fixed k* the curves in Fig. 2a are almost *equidistant*. Hence, we conclude that for the $n$ th order nonlinear transmission resonance approximately holds $V_1 = n(V_{NL1} - V_{L1})$, so that

$$V_{NLn} = V_{Ln} + \frac{g\,n}{1 + 2(\mu - g)}. \tag{29}$$

The numerical testing shows that this is an already good approximation. The formula offers a quick estimation of the resonance position with relative error of the order of a few percents for the whole variation range of the involved chemical potential $\mu$ provided the nonlinearity parameter $g$ is small, $g/\mu \leq 0.25$. Being of remarkably simple structure, the formula may be useful for practical purposes as well as for qualitative considerations.

To proceed further, we note that the curves shown in Fig. 2a suggest that the whole variation range of the wave vector $k$ can conventionally be divided into two interaction regions. The first region is the vicinity of the origin $k = 0$, where a negative second derivative $d^2V_1/d^2k$ is observed. The second region corresponds to large $k$, $k > 0.5$, where the mentioned second derivative is positive.

Since the vicinity of the point $k = 0$, i.e., the case of small $k$, is of particular interest because the most intensive interaction of the matter-wave with the potential well occurs namely in this case we first study the point $k = 0$. Note that in this case the situation is considerably simplified since the pre-factor under the integral in Eq. (23) disappears, the constant $C_n$ is reduced to $C_n = 1/(2n+1)$, and the eigenfunctions of the linear problem $u_{Ln}$ become real and are simplified to the Legendre polynomials $P_n(1-2z)$ [16]. Then, using the known asymptotes of these functions [16] we arrive at the following approximation

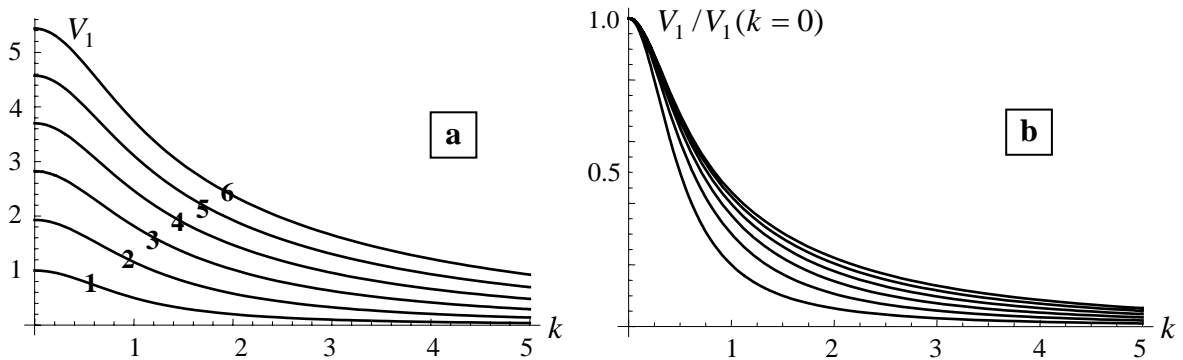

Fig. 2. a) The nonlinear shift of the resonance position $V_1$ vs. the wave vector $k$. The resonance orders are indicated by bold-face numbers. b) The same curves as in figure a) normalized to the unity at $k = 0$.

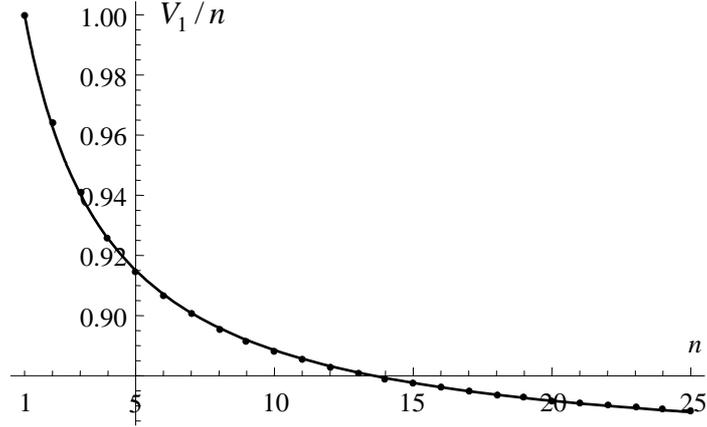

Fig. 3. Deviation from the linear law (23) of the dependence of the shift of the resonance position on the resonance order $n$ in the case $k = 0$. Circles indicate the exact result, the solid line presents formula (30).

$$V_{NLn}(k=0) = V_{Ln} + g\,n\left(1 - \frac{\sqrt{\pi}}{12}\frac{n-1}{n+2}\right). \tag{30}$$

The comparison of this formula with the exact result is shown in Fig. 3. As it is seen, the formula is highly accurate for all the resonance orders $n$.

Having at hand the formula for $V_{NLn}(k=0)$ we now turn to the behavior of $V_1(k,n)$ for $k > 0$. To proceed, we examine the family of *normalized* curves $V_1/V_1(k=0)$ shown in Fig. 2b. Numerical simulations show that this family can be very accurately described using a Gauss hypergeometric function ${}_2F_1$. However, a different guess is that the curves can approximately be constructed by a similarity transformation starting from the curve describing the first resonance $n = 1$. The numerical simulations towards verification of this conjecture then reveal that the family can be well approximated as the power

$$\frac{V_1(k,n)}{V_1(k=0)} = \left(\frac{1}{1+k^2}\right)^{1+a}, \tag{31}$$

where $a$ adopts values of the order of 1 for all $k$ and $n$. It is understood that in order to produce the exact result (25) for the first resonance this constant should vanish at $n = 1$. A simple appropriate approximation for $a$ is readily established by numerical fitting:

$$a = -b(k)\frac{n-1}{n+1}, \tag{32}$$

where the coefficient $b(k)$ is a slightly varying function of $k$ if the best fit is considered (an example of such an accurate fit with varying $b(k)$ is shown in Fig. 4). However, for all the resonance orders the variation range of $b$ is restricted within the narrow interval $[0.5, 0.75]$ as $k$ is varied from zero to infinity. Besides, this variation does not affect much the resultant value of $V_1$ (the absolute change is of the order of $10^{-2}$), hence, we simply put $2/3$ for the coefficient as an average value. The derived approximation is then checked by analytic methods using a limit function as an appropriate approximation for $p = |u_{Ln}|^2$ constructed for higher order resonances using an exact third-order equation for the probability $p$.

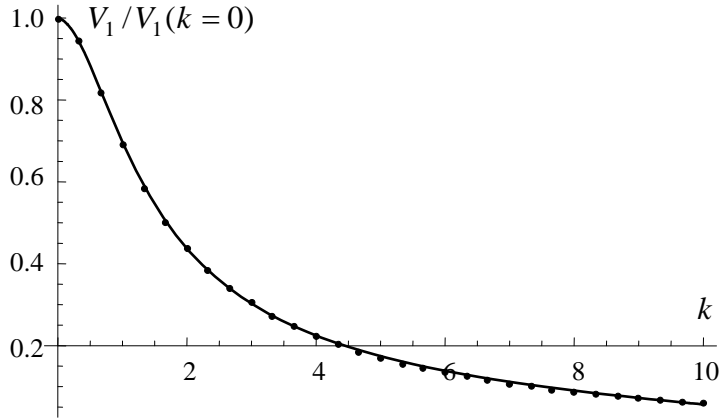

Fig. 4. Fitting the exact result for integral (23) for the sixth transmission resonance $n = 6$ using formulas (31) and (32). Circles indicate the exact result, the solid line presents the formulas. The coefficient $b(k)$ in Eq. (32) is slowly varied within the interval $[0.5, 0.75]$ as $k$ is varied from zero to infinity.

Collecting the presented developments, we thus obtain the following formula:

$$V_{NLn} = V_{Ln} + g\, n \left(1 - \frac{\sqrt{\pi}}{12} \frac{n-1}{n+2}\right) \left(\frac{1}{1 + 2(\mu - g)}\right)^{1 - \frac{2}{3} \frac{n-1}{n+1}}. \tag{33}$$

This is a fairly good approximation. It accurately determines the position of the nonlinear transmission resonance for all $k$ and $n$; the relative error does not exceed $10^{-3}$ if $g/\mu \leq 0.25$ and it remains less then one percent up to $g/\mu = 0.5$.

To summarize, we have discussed, within the mean field Gross-Pitaevskii approximation, the quantum above-barrier reflection of ultra-cold atoms by the squared hyperbolic-secant Rosen-Morse potential. We have shown that the problem of reflectionless transmission can be reformulated as a quasi-linear eigenvalue problem for the potential depth. Applying then a modified variant of the Rayleigh-Schrödinger time-independent perturbation theory, we have derived an approximation for the specific height of the potential that supports reflectionless transmission (as a result of the common action of the potential and the nonlinearity) of the incoming matter wave. The approximation provides highly accurate description of the resonance position for all the resonance orders if the nonlinearity parameter is small compared with the chemical potential. However, notably, the result for the first transmission resonance is exact, i.e., the derived formula for the resonant potential height gives the exact value of the first nonlinear resonance's position for all the allowed variation range of the involved parameters, i.e., the nonlinearity parameter and the incoming particle's wave number. This has been shown by constructing the exact solution of the problem for the first resonance. The constructed approximation reveals that, in contrast to the linear case, in the nonlinear case reflectionless transmission may occur not only for potential wells but also for potential barriers. Finally, we have constructed, via combination of analytical and numerical methods, a compact (yet, highly accurate) analytic approximation for the $n$ th order resonance position. We have seen that the nonlinear shift of the resonance position is sketched as a linear function of the resonance order.

**Acknowledgments**

This work was supported by the Armenian National Science and Education Fund (ANSEF Grant No. 2009-PS-1692).